\documentclass[aps,prl,twocolumn,groupedaddress,floats,showpacs]{revtex4}
\usepackage{graphicx}
\usepackage{bm}
\begin{document}
\def\he4{$^4$He}
\def\Am2{\AA$^{-2}$}
\def\beq{\begin{equation}}
\def\eeq{\end{equation}}
\title{Worm Algorithm for Continuous-Space Path Integral Monte Carlo Simulations}
\author{Massimo Boninsegni$^{1}$, Nikolay Prokof'ev$^{1-4}$, and Boris Svistunov$^{2,3}$}
\affiliation{ ${^1}$Department of Physics, University of Alberta,
Edmonton, Alberta T6G 2J1\\
${^2}$Department of Physics, University of
Massachusetts, Amherst, MA 01003 \\
${^3}$Russian Research Center ``Kurchatov Institute'', 123182 Moscow
\\  ${^4}$Department of Physics, Cornell University, Ithaca, NY
14850
}

\begin{abstract}
We present a new approach to path integral Monte Carlo (PIMC)
simulations based on the worm algorithm, originally developed for
lattice models and extended here to continuous-space many-body
systems. The scheme allows for efficient computation of 
thermodynamic properties, including winding numbers and
off-diagonal correlations, for systems of much greater size than
that accessible to conventional PIMC. As an illustrative
application of the method, we simulate the superfluid transition
of \he4 in two dimensions.
\end{abstract}

\maketitle 
Over  the past two decades, PIMC simulations have played a major role in the
theoretical  investigation of quantum many-body
systems, not only by providing reliable quantitative results, 
but also by  shaping our current conceptual
understanding, e.g., of the relationship between superfluidity and
Bose condensation. At least for Bose systems, PIMC is the
{\it only} presently known method capable of furnishing in principle {\it
exact} numerical estimates of physical quantities, including the
superfluid density, and the condensate fraction\cite{ceperley95}.

As PIMC is currently the most realistic option to investigate ever
more complex quantum many-body systems, the issue arises of
overcoming its present limitations. Aside from the notorious {\it
sign} problem, the main bottleneck of the current PIMC technology is
inarguably the maximum system size (i.e., number $N$ of particles)
for which accurate estimates can be obtained, in a reasonable amount
of computer time. Almost two decades since the pioneering work of
Pollock and Ceperley\cite{pollock87}, who simulated the superfluid
transition in bulk liquid \he4 on a system of $N$=64 atoms, no
further advance has been made, despite a hundredfold increase in
computer speed\cite{pt}.

For such quantities as energy and diagonal correlations, one can
often approach the thermodynamic limit ($N \to \infty$), by studying
systems comprising as few as $\sim$ 30 particles. However, accurate
predictions of superfluid properties of liquids (and solids
\cite{chan}!) require that the superfluid and condensate fractions,
$\rho_S$ and $n_\circ$, be computed  for large systems of
significantly different sizes. In conventional PIMC, $\rho_S$ is
obtained by means of the so-called winding number
estimator\cite{pollock87}, which can only take on a nonzero value if
long permutation cycles of identical particles occur in the system.
Because the sampling frequency for such cycles decreases
exponentially with $N$, ensuring the ergodicity of the algorithm becomes
problematic\cite{bernu04}.

This hurdle seems difficult to conquer within any scheme formulated
in the canonical ensemble, in which the winding number is
``topologically locked'' in the $N\to \infty$
limit\cite{ceperley95}. On the other hand, the same hurdle has been
completely overcome in quantum Monte Carlo simulations of lattice
models. A lattice Path Integral scheme based on an alternative
sampling approach, known as {\it worm algorithm} (WA) \cite{worm},
allows for efficient calculations of winding numbers and
one-particle Green function $G$, for systems of as many as $\sim$
10$^6$ particles\cite{prokofev04}. 
A fundamental aspect of the WA is that
it operates in an extended configurational space, containing both
closed world-line configurations (henceforth referred to as $Z$- or 
diagonal configurations), contributing to the partition function $Z$, 
as well as configurations containing one open line (worm). The latter
configurations contribute to the one-particle Green function; below, they 
are referred to as $G$- (or, off-diagonal) configurations. All topologically
non-trivial modifications of world lines occur in the off-diagonal
configurational space, where there are no constraints; when the
sampling process generates a diagonal configuration, the number of
particles and the winding number are updated.

In this Letter, we describe the extension of the WA to the PIMC
simulation of quantum many-body systems in continuous space. Our
novel PIMC implementation, while based on the same theoretical
underpinnings\cite{feynman49}, {\it differs fundamentally} from the
``canonical" one\cite{ceperley95}, both in the configuration space
structure, as well as in the sampling method.  Since the number of
continuous configuration variables is no longer conserved, the new
scheme necessarily belongs to the generic domain of diagrammatic
Monte Carlo methods\cite{DMC}. 
As an illustrative application of this method, we
simulate the superfluid transition in liquid $^4$He in two
dimensions (2D), for systems with up to $N$=2500 particles, i.e. two
orders of magnitude larger than in the most recent PIMC study\cite{gordillo98}.
In particular, we observe a dramatic speed-up in
convergence of $\rho_s$ and $G$.

We begin by reviewing conventional PIMC. One obtains averages of
physical quantities (at a temperature $T$) over a set of many-particle configurations
$\{R\}$, statistically sampled from a probability density
proportional to $\rho(R,R,\beta)\equiv\langle R|e^{-\beta\hat
H}|R\rangle$, where $\beta=1/T$ (we set $k_B$=1) and $\hat H$ is the system
Hamiltonian. The goal is achieved by sampling discrete many-particle
paths $X\equiv (R_1, R_2, \ldots , R_P)$, periodic in the imaginary
time interval $\beta=P\tau$, from the probability density
\beq\label{start} \rho(X)\, =\, {\rm e}^{-U(X)} \prod_{j=1}^P
\rho_\circ(R_j,R_{j+1},\tau) \eeq
where $\rho_\circ(R_j,R_{j+1},\tau)=\prod_{i=1}^N\rho_\circ({\bf
r}_{ij},{\bf r}_{i,j+1},\tau)$ is a product of $N$ free-particle
propagators, whereas $U$ incorporates correlations, both in space
and in imaginary time, arising from interactions among particles.
$U$ is chosen so that, in the $\tau\to 0$ limit, the distribution of configurations
$R$ visited by 
paths $X$ reproduce $\rho(R,R,\beta)$. Several choices are possible
\cite{ceperley95} for $U$, but our algorithm does not depend on its
particular form.

Eq.~(\ref{start}) implies the following configuration space
structure: $N$ single-particle paths (world lines), labeled
$i=1,2,\ldots ,N$, propagating in the discretized imaginary  time
$t$ from $t_1=0$ to $t_{P+1}=\beta$. Each world line is formed by
$P$ successively linked ``beads" labeled by the number of the
corresponding time slices $j=1,2,\ldots ,P$. The $j$-th bead of the
$i$-th world line is positioned at ${\bf r}_{ij}$. The
$\beta$-periodicity implies that the $(P+1)$-th bead of each world
line coincides with the first bead of either the same, or another
world line.

The set of  paths $\{X_l \}$ is sampled by a Metropolis random
walk through configuration space. In order to generate $X_{l+1}$
from the current path $X_l$, a local space-time modification of
$X_l$ is proposed. The new path $X^\star$ is then either accepted,
$X_{l+1}\equiv X^\star$, or rejected, $X_{l+1}\equiv X_l$, based
on (\ref{start}), according to the standard procedure
\cite{metropolis}.

The basic update $X_{l}\to X^\star$ consists of deforming one or
more randomly selected world lines, over a number 1$\le m \le P$ of
successive links. In order to incorporate effects of quantum
statistics, it is also crucial to allow groups of $1 <n\le N$ world
lines to exchange: A modified portion of a world line in the group
will connect, $m$ links later, to a different world line, among the
$n$ selected. Note that the number of world lines along $X$ is
always equal to $N$, in this scheme. Updates with arbitrary exchange
cycles ensure ergodicity of the algorithm. Typically, however, the
acceptance rate for 
permutations is frustratingly low, particularly
in the presence of repulsive inter-particle potentials (e.g., in
condensed helium), rendering the calculation inefficient, and impractical for large
$N$.

The WA described here has the same starting point, namely
Eq.~(\ref{start}), but with a crucial generalization of the
configuration space, which now includes both the above-mentioned
diagonal and off-diagonal paths, the latter corresponding to the
representation (analogous to Eq.~(\ref{start})) of the one-particle
Matsubara Green function $G({\bf r}, t)$. Each off-diagonal
configuration contains a {\it worm}, that is, a world line (on a $\beta$-cylinder)
with two ends---the ``head" and the ``tail"---
corresponding to the Green function annihilation and creation
operators, respectively. The two special beads at the open world
line ends are named (for historical reasons) {\it Ira} (${\cal I}$)
and {\it Masha} (${\cal M}$). Configurations in which ${\cal I}$ and
${\cal M}$ are located in space-time at points $({\bf r}_{\cal I},
t_{\cal I})$ and $({\bf r}_{\cal M}, t_{\cal M})$ contribute to
$G({\bf r}_{\cal I}-{\bf r}_{\cal M}, t_{\cal I}-t_{\cal M})$ with
the weight defined in accordance with the generalized
Eq.~(\ref{start}).

The sampling of paths $\{X_l \}$ is implemented in WA {\it
exclusively} through a set of simple, local updates evolving ${\cal
I}$  (or, ${\cal M}$) in space-time. The particle number becomes
configuration- and time-dependent (there is one less particle
between ${\cal I}$ and ${\cal M}$, than in the rest of the path). In
other words, the WA opens up the possibility to work in the {\it
grand canonical} ensemble, with the chemical potential $\mu$ being
an input parameter\cite{note2}.

\begin{figure}[h]
\centerline{\includegraphics[angle=-0,width=3.4in]{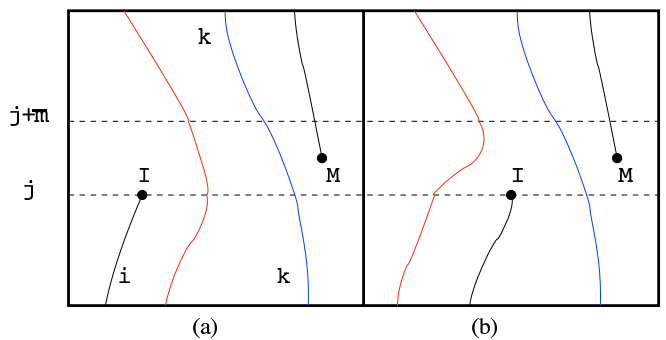}}
\caption{(Color online). Schematic illustration of {\it Swap} move
described in the text. {\it (a)}: before the move. {\it (b)}:
after the move. } \label{f1}
\end{figure}

Next, we describe the set of ergodic
local updates which sample our extended configuration space,
switching between the $Z$- and $G$-sectors. Updates which change the
number of continuous variables in $X$, are arranged in complementary
pairs, satisfying detailed balance. General principles of balancing
complimentary pairs can be found in Ref.~\cite{DMC}. We have three
pairs: {\it Open/Close}, {\it Insert/Remove}, and {\it
Advance/Recede}. Only the {\it Swap} update in the list below does
not fall in this category, because it preserves the number of
variables, i.e., it is self-complementary. 
Naturally, all known standard tricks can be used, in order to 
enhance performance.
\\
\indent  {\it (1a) Open}. This update is only possible if the
configuration is diagonal.\cite{note3} A world line (say the
$i$-th) and a bead (say the $j$-th) are selected at random. A
random number $(m-1)$ of beads, namely $j+1$, $j+2$, $\ldots$,
$j+m-1$ are removed, so that a worm appears with  ${\cal I}$ at
$({\bf r}_{ij}, t_j)$ and ${\cal M}$ at $({\bf r}_{i,j+m},
t_{j+m})$. Hence, the difference between the proposed new path,
$X^\star$, and the previous one, $X$, is that instead of the $i$-th
world line there are now two new ones: The $i$-th world line (we
retain the same label) now ends at the $j$-th bead (${\cal I}$) and
the $i_0$-th world line (we introduce a new label) corresponds to
the piece of the original $i$-th world line starting from the
$(j+m)$-th bead (${\cal M}$). The acceptance probability for this
update is
\beq P_{\rm op} = {\rm min} \biggl \{ 1,\: \frac{CN_XP \:\bar{m}\;
{\rm e}^{\Delta U  -\mu m\tau }} {\rho_\circ({\bf r}_{ij},{\bf
r}_{i,j+m},m\tau)}\:
 \biggr \}\; , \eeq
where $\Delta U = U(X)-U(X^\star)$,  $N_X$ is the number of world
lines (particles) in the diagonal configuration $X$, and an
arbitrary constant $C$ controls the relative statistics of
$Z$- and $G$-sectors. The number $\bar{m}<P$ defines the interval
for $m$: $m \in [1,\bar{m}]$. In practice $\bar{m}$ is adjusted to
ensure the desired acceptance rate. Due to the $\beta$-periodicity,
without loss of generality we assume that whenever the situation
$j+m > P$ occurs, the enumeration is shifted in such a way that
$j+m\le P$, and no ambiguity occurs. These definitions are common to
all other moves described below, in which $m$ and $\bar m$ enter.
\\
\indent {\it (1b) Close}. This update is only possible if the
configuration is off-diagonal.\cite{note3} Let ${\cal I}$ be the
$j$-th bead of the $i$-th worldline and  ${\cal M}$ be the
$(j+m)$-th bead of the $i_0$-th worldline. If $m > \bar{m}$, the
move is rejected. If $m\le \bar{m}$, one proposes to generate a
piece of world line connecting  ${\cal I}$ to ${\cal M}$, thereby
rendering the configuration diagonal.
The corresponding spatial
positions of new $(m-1)$ beads, ${\bf r}_{i,j+1},\ldots ,{\bf
r}_{i,j+m-1}$, are sampled from the product of $m$ free-particle
propagators $\prod_{\nu=1}^{m} \rho_\circ({\bf r}_{i, j+\nu -1},{\bf
r}_{i,j+\nu},\tau)$. The probability to accept the move is
\beq P_{\rm cl} = {\rm min} \biggl \{ 1, \: \frac{ \rho_\circ({\bf
r}_{ij},{\bf r}_{i,j+m},m\tau)\, {\rm e}^{\Delta U +\mu m\tau
}}{CN_{X^\star} P\: \bar{m} } \biggr \}  \; . \eeq
If the move is accepted, the label $i_0$ is removed.
\\
\indent {\it (2a) Insert}. The other way to create an off-diagonal
configuration from a diagonal one is to seed a new $m$-link long
open world line in vacuum.  The number of links $m\leq \bar{m}$ and
the position of ${\cal M}$ in space-time are selected at random. The
spatial positions of the other $m$ beads are generated from the
product of $m$ free-particle propagators. The move is accepted with
probability $P_{\rm in} = {\rm min} \{ 1,\: CVP \:\bar{m}\, {\rm
e}^{\Delta U +\mu m\tau } \}$, where $V$ is the system volume.
\\
\indent {\it (2b) Remove.} The removal of the worm, i.e., the world line connecting
${\cal M}$ to ${\cal I}$, is attempted if its length
(in the $\beta$-periodic sense)  is $m \leq
\bar{m}$. (If $m > \bar{m}$, the proposal is rejected \cite{note3}.)
The acceptance probability for the move is $P_{\rm rm} = {\rm min}
\{ 1, {\rm e}^{\Delta U-\mu m \tau}/CVP \:\bar{m} \}$. We are in
position now to select $C$. A natural choice would be
$C=1/VP\:\bar{m}$,  so that the probability to {\it Open} a worm of
zero length is $N/V \equiv G(0,-0)$.
\\
\indent {\it (3a) Advance}. This move advances ${\cal I}$ a random
number $m$ of slices forward in time. It is similar to {\it Insert}
update in implementation. The acceptance probability is $P_{\rm ad}
= {\rm min} \{ 1, {\rm e}^{\Delta U + \mu m\tau} \}$. Note that it
is possible for ${\cal I}$ to advance past ${\cal M}$.
\\
\indent {\it (3b) Recede}. Now ${\cal I}$ moves backwards in time
(in the $\beta$-periodic sense) by erasing $m$ consecutive links,
the number $1\le m \le \bar{m}$ is selected at random. The
acceptance rate is $P_{\rm re} = {\rm min} \{ 1, {\rm e}^{\Delta U -
\mu m\tau} \}$. If $m$ turns out to be equal or larger than the
total number of links between ${\cal I}$ and ${\cal M}$ along the
world line connecting them, the update is  rejected \cite{note3}.
\\
\indent {\it (4) Swap.} Let ${\cal I}$ be positioned on the $i$-th
world line at the $j$-th time slice. (See Fig.~\ref{f1}.) Consider
all the world lines intersecting the $(j+\bar{m})$-th (in the
$\beta$-periodic sense) time slice and select one of them (labeled
below with $k$) with the probability $T_k=\rho_\circ({\bf r}_{i
j},{\bf r}_{k, j+\bar{m}},\bar{m} \tau)/ \Sigma_i $ where
\beq \Sigma_i= \sum_{l}\rho_\circ({\bf r}_{i j},{\bf
r}_{l,j+\bar{m}},\bar{m} \tau )\;
\eeq 
is the normalization factor (if the selected world
line contains ${\cal M}$ at $j'$-th time slice, such that $j' \in [j
, j+\bar{m}]$, the move is rejected). A set of random
positions ${\bf r}_{i, j+1}, \ldots , {\bf r}_{i, j+m-1}$
 is then generated as in the {\it Close} move, whereas the beads
 ${\bf r}_{k,j+1}, \ldots , {\bf r}_{k, j+m-1}$ are all erased.  ${\cal I}$
 is shifted to ${\bf r}_{k j}$, while the world line $i$ reconnects with the rest of
 the world line $k$, which implies re-labeling,
 as illustrated in Fig.~\ref{f1}.
 The move is accepted with probability
 $P_{\rm sw} = {\rm min} \{ 1, {\rm e}^{\Delta U} \Sigma_i /\Sigma_k  \}$.

The {\it Swap} move generates all possible many-body permutations
through a chain of local single-particle updates. Since no two
particles need be brought within a distance of the order of the
potential hard core, it enjoys a high acceptance rate, similar to
that for the {\it Advance/Recede} procedures. It must be
emphasized that in our algorithm, unlike in conventional PIMC,
arbitrary permutations of identical  particles, as well as
macroscopic exchange cycles {\it appear automatically}, if the
physical conditions warrant them. 
This is because the statistics of
the relative positions for the worm ends is given exactly by the
Green function $G(r,t)$.

\begin{figure}[t]
\centerline{\includegraphics[angle=-90, width=3.6in]{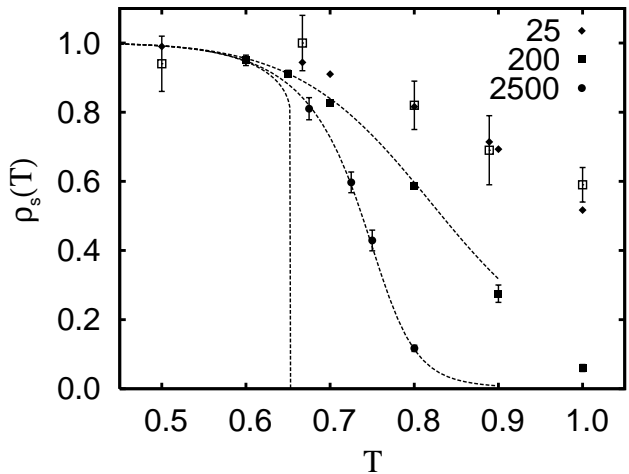}}
\caption{Superfluid fraction $\rho_S(T)$ computed for
2D \he4 on systems with different numbers $N$ of \he4 atoms. The system
density is $\rho=0.0432$ \Am2. Dashed lines represent fits to the
 numerical data (in the critical region) obtained using the procedure
 illustrated in Ref.~\cite{ceperley89}. The leftmost dashed line is the
 extrapolation to the infinite system. Open squares show results obtained
 in Ref.~\cite{ceperley89} for the same system, with $N$=25. }
\label{f2}
\end{figure}

 As an illustrative application of the new method, we present
here simulation results of the superfluid transition in 2D helium.
Specifically, we have repeated the study first carried out in
Ref.~\cite{ceperley89}, using the same interatomic potential
\cite{aziz79} and at the same 2D density $\rho=0.0432$ \AA$^{-2}$, 
    but on systems with  a number of
particles up to hundred times larger. We used an 
approximation accurate up to $\tau^4$ \cite{chin} for the
high-temperature  density matrix [which determines the structure of
the function $U(X)$], and extrapolated the results to the $\tau \to
0$ limit. For a given choice of $\tau$, 
the statistical error of $\rho_s$ is comparable to that of the kinetic energy.

Fig.~\ref{f2} shows our results for the superfluid fraction
$\rho_S(T)$, obtained on systems comprising different numbers
$N$ of atoms. Using the procedure illustrated in
Ref.~\cite{ceperley89}, based on Kosterlitz-Thouless
theory\cite{kt78}, we have obtained numerical fits to our data, in
the critical temperature range 0.65 $K$ $\le T \le$ 0.8 $K$. Our
estimates for the values of the fitting parameter are $d=8.8\pm
0.5$ \AA\ for the vortex core diameter, and $E$=2.18 $\pm$ 0.04 K
for the vortex energy, which lead to an estimate for the critical
temperature $T_{\rm c}$=0.653$\pm$0.010 K, significantly different
from the previous result, 0.72$\pm$0.02 K, deduced from the $N=25$
data\cite{ceperley89}.

\begin{figure}[t]
\centerline{\includegraphics[angle=-90, width=3.3in]{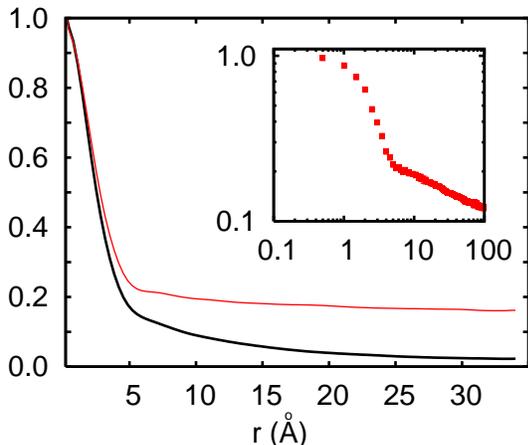}}
\caption{(Color online). One-particle density matrix computed for 2D \he4
at a density $\rho=0.0432$ \Am2 for a system of 200 atoms, at $T$=0.675 K
(upper curve) and $T$=1.0 K (lower curve). Statistical errors on the curves 
are very small, and not shown for clarity. 
In the inset we present data
 (on a log-log scale) for the N=2500 system at $T$=0.675 K, with clear
 signatures of the Kosterlitz-Thouless behavior in the vicinity of the critical point.
  }
\label{f3}
\end{figure}

As mentioned above, our method gives easy access to the
imaginary-time one-particle Green function, and therefore to the
one-body density matrix. For 2D helium, this quantity  is expected
to decay to zero at all temperatures, following a slow power-law
behavior for $T \le T_c$.  Typical results obtained in
this study, for systems with $N$=200 and $N=2500$ are shown in Fig.
\ref{f3}. 

In conclusion, we have implemented a novel procedure to perform
large-scale PIMC simulations. Our scheme extends to continuous
space the worm algorithm previously developed for lattice systems,
and affords efficient computations of 
thermodynamic properties, including the superfluid density and the
single-particle Green function, for system of significantly larger
size than accessible to the existing PIMC technology.

This work was supported by the National Aero and Space
Administration grant NAG3-2870, the National Science Foundation
under Grants Nos. PHY-0426881, NSF PHY-0456261, by the Sloan
Foundation, and by the Natural Science
and Engineering Research Council of Canada under grant G121210893.


\begin{thebibliography}{99}
\bibitem{note2} Using standard importance sampling tricks one may
control statistics of particle number fluctuations, and perform
canonical ensemble measurements.
\bibitem{note3} Since balance equation deals only with {\it
accepted} moves, one can propose a move without {\it a priori}
checking that the current configuration is consistent with the
proposal. If it is not, the move is rejected with probability
one.
\end{thebibliography}
\end{document}